\documentclass[12pt,a4paper,dvips]{article}
\usepackage{a4p}
\usepackage{epsfig}
\usepackage{cite}
\usepackage{graphicx}
\usepackage{physics}
\usepackage{l3_titlePH,ifthen,Lep}

\date{February 23, 2005}

\preprint{2005-007}

\newlength{\capindent}
\newlength{\capwidth}
\newlength{\figwidth}
\newcommand{\icaption}[2][!*!,!]{\hspace*{\capindent}%
  \begin{minipage}{\capwidth}
    \ifthenelse{\equal{#1}{!*!,!}}%
      {\caption{#2}}%
      {\caption[#1]{#2}}
  \end{minipage}}

\def\PLB{{Phys. Lett.} {\bf B }}

\def\PRD{{Phys. Rev.} {\bf D }}
\def\ZPC{{Z. Phys.} {\bf C }}
\def\CPC{Comp. Phys. Comm. }

\def\ra{\rightarrow}

\def\be{\begin{equation}}
\def\ee{\end{equation}}
\def\bea{\begin{eqnarray}}
\def\eea{\end{eqnarray}}

\begin{document}
\begin{titlepage}
\title{Measurement of the\\ Cross Section for Open-Beauty Production\\ 
in Photon-Photon Collisions at LEP}
\author{The L3 Collaboration}
%
% The abstract
%

\begin{abstract}

The cross section for open-beauty production in photon-photon collisions
is measured using the whole high-energy and high-luminosity data
sample collected by the L3 detector at LEP. This corresponds to 627
$\mathrm{pb^{-1}}$ of integrated luminosity for electron-positron
centre-of-mass energies from $189 \GeV$ to 209 \GeV.  Events containing b
quarks are identified through their semi-leptonic decay into electrons
or muons. The $\rm e^+e^-\rightarrow e^+e^-b \bar b X$ cross section
is measured within our fiducial volume and then extrapolated to the full
phase space. These results are found to be in significant excess
with respect to Monte Carlo predictions and next-to-leading
order  QCD calculations. 

\end{abstract}

\submitted

\end{titlepage}

%%%%%%%%%%%%%%%%%%%%%%%%%%%%%%%%%%%%%%%%%%%%%%%%%%%%%%%%%%%%%%%%%%%%%%%%%%%%%
\section{Introduction}
%%%%%%%%%%%%%%%%%%%%%%%%%%%%%%%%%%%%%%%%%%%%%%%%%%%%%%%%%%%%%%%%%%%%%%%%%%%%%

The production of b quarks through hard processes constitutes a unique
environment for the study of perturbative QCD, as the mass of the b
quark, $m_{\rm b}$, largely exceeds the typical non-perturbative scale
of hadronic interactions. High-energy hadron colliders are copious
sources of b quarks and therefore extensive experimental studies and
QCD calculations have been performed. Much debate has taken place on the
apparent disagreement between the measured cross section for b-quark
production in p$\bar{{\rm p}}$ collisions at the
Tevatron~\cite{tevatron} and the next-to-leading order (NLO) QCD
calculations~\cite{tevatronQCD}. The first measurements of open beauty
production in $\rm e^\pm p$ collisions at HERA were found to be
markedly higher than NLO QCD predictions~\cite{HERA12}.  Some more
recent measurements were in better agreement~\cite{HERA34}, while
others still showed an excess\cite{HERA5,HERA6}. A comparison of these
different measurements with NLO QCD predictions is shown in
Reference~\citen{HERA6}.

\begin{figure}[htbp]
  \begin{center}
    \mbox{\epsfig{file=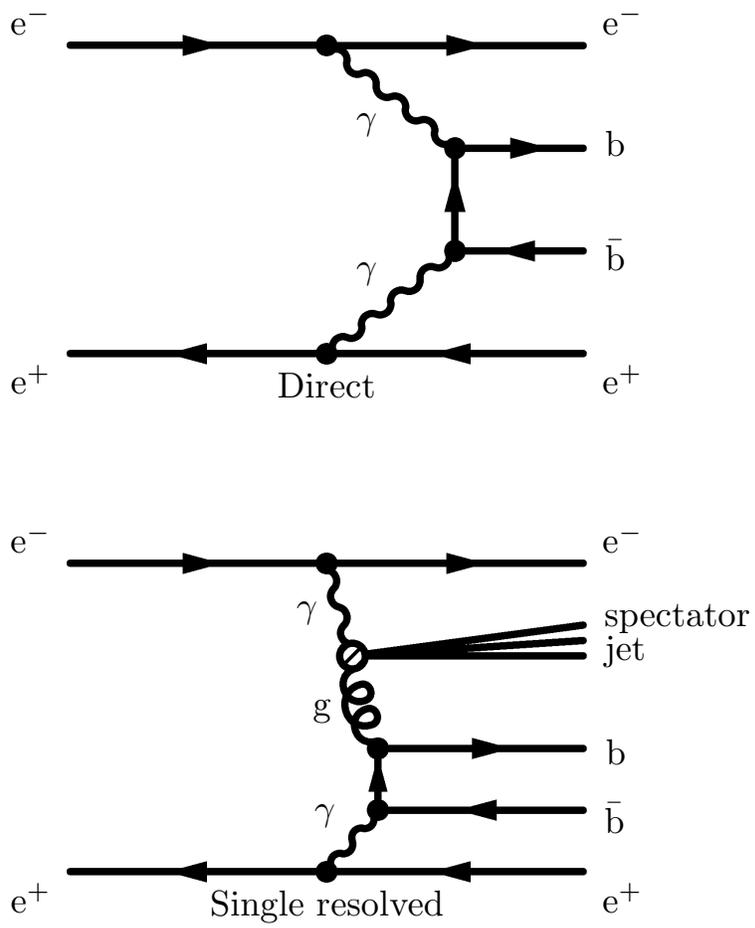, height=.5\textheight}}  
    \caption{Dominant diagrams contributing to open-beauty
      production in photon-photon
      collisions at LEP.}
    \label{fig:Feynman}
  \end{center}
\end{figure}

Photon-photon collisions at $\epem$ colliders also give access to the
hard production of b quarks. The LEP $\epem$ centre-of-mass energy,
$\sqrt{s}$, was around $200\GeV$. In this environment b quarks are
expected to be produced with comparable rates by the direct and
single-resolved processes~\cite{theory}, illustrated in
Figure~\ref{fig:Feynman}.  The main contribution to the
resolved-photon cross section is the photon-gluon fusion process. The
rates of both the direct and the single-resolved process depend on
$m_{\rm b}$, while the latter also depends on the gluon density in the
photon.

The first measurement of the cross section for the
$\rm\epem\rightarrow\epem b \bar{b} X$ process was published by the L3
collaboration using 410 pb$^{-1}$ of data collected at
$\sqrt{s}=189-202\GeV$~\cite{L3bb}. The results were found to be in
excess of the QCD prediction by a factor of three. Since these first
findings, compatible preliminary results were obtained by other LEP
collaborations~\cite{otherLEP}. In this Letter, we extend our
measurement to the whole high-energy and high-luminosity data sample
collected at LEP with the L3 detector~\cite{L3det}, corresponding to
627 pb$^{-1}$ at $\sqrt{s}=189-209\GeV$.

Hadronic events from photon-photon interactions are selected through
their specific multiplicity and topology. The production of b quarks
is then tagged by the detection of electrons\footnote{Throughout this
letter, the term `electron` stands for both electrons and positrons.}
or muons from their semi-leptonic decays. The cross section of the
$\rm\epem\rightarrow\epem b \bar{b} X$ process is measured in a phase
space which reflects the energy thresholds used in the analyses and
the fiducial volume for lepton identification: the lepton momentum
must exceed $2\GeV$ and the angle, $\theta$, between the leptons and the
beam line must satisfy $|\cos\theta|<0.725$ for electrons and
$|\cos\theta|<0.8$ for muons, respectively. For the first time the
experimental results are compared to Monte Carlo predictions in this
fiducial volume. An extrapolation factor is then applied to compare
the measured cross section with the QCD predictions in the full phase
space.

%%%%%%%%%%%%%%%%%%%%%%%%%%%%%%%%%%%%%%%%%%%%%%%%%%%%%%%%%%%%%%%%%%%%%%%%%%%%%
\section{Monte Carlo Simulations}
%%%%%%%%%%%%%%%%%%%%%%%%%%%%%%%%%%%%%%%%%%%%%%%%%%%%%%%%%%%%%%%%%%%%%%%%%%%%%

The PYTHIA~\cite{pythia} Monte Carlo generator is used to model hadron
production in photon-photon collisions. Final states without b quarks
are generated with massless matrix elements~\cite{massless} while
massive matrix elements are used for b-quark production.  Resolved
processes are described by means of the SaS1d parton density
function~\cite{sas1d}.  The photon-photon luminosity function is
implemented in the equivalent photon approximation~\cite{Budnev} with
a cutoff for the virtuality of the interacting photons $\mathrm{Q^2 <
m_{\rho}^2}$ .

Potential backgrounds are simulated by the following Monte
Carlo generators: JAMVG~\cite{verm} for the $\mathrm{e^{+}e^{-} \ra
e^{+}e^{-} \tau^{+} \tau^{-}}$ process, PYTHIA for $\mathrm{e^{+}e^{-}
\rightarrow q \bar{q}}$, KORALZ~\cite{koralz} for $\mathrm{e^{+}e^{-}
\ra \tau^{+} \tau^{-}}$ and KORALW~\cite{koralw} for
$\mathrm{e^{+}e^{-}} \ra\mathrm{W^{+} W^{-}}$.

The L3 detector is simulated using the GEANT~\cite{GEANT} and
GHEISHA~\cite{GHEISHA} packages.  Monte Carlo events are then
reconstructed in the same way as the data.  Time-dependent detector
inefficiencies, as monitored during the data-taking period, are
included in the simulations.

%%%%%%%%%%%%%%%%%%%%%%%%%%%%%%%%%%%%%%%%%%%%%%%%%%%%%%%%%%%%%%%%%%%%%%%%%%%%%
\section{Event Selection}
%%%%%%%%%%%%%%%%%%%%%%%%%%%%%%%%%%%%%%%%%%%%%%%%%%%%%%%%%%%%%%%%%%%%%%%%%%%%%

The selection of events originating from the $\rm\epem\rightarrow\epem
b \bar{b} X$ process is unchanged with respect to
Reference~\citen{L3bb}.  Hadrons produced in photon-photon collisions
are selected by means of three criteria. First, at least five charged
tracks are required, thus suppressing background from the
$\mathrm{e^{+}e^{-} \ra e^{+}e^{-} \tau^{+} \tau^{-}}$ and
$\mathrm{e^{+}e^{-} \ra \tau^{+} \tau^{-}}$ processes. Second, the
visible energy of the event, $E_\mathrm{vis}$, is required to satisfy
$E_\mathrm{vis}<\sqrt{s} / 3$, in order to reject events from the
$\mathrm{e^{+}e^{-} \ra q \bar{q}}$ annihilation process and further
eliminate events from the $\mathrm{e^{+}e^{-} \ra \tau^{+} \tau^{-}}$
process. Finally, possible instrumental background and uncertainties
in the trigger procedure are reduced by requiring the event visible
mass, $W_\mathrm{vis}$, to satisfy $W_\mathrm{vis} > 3 \GeV$.
$W_\mathrm{vis}$ is calculated from the four momenta of reconstructed
tracks and of isolated calorimetric clusters. In this calculation, the
pion mass is associated to the tracks while the clusters are treated
as massless. Clusters in the low-angle luminosity monitor are included
in this calculation.

In addition to these cuts, the analysis is restricted to events with
small photon virtuality by removing events with clusters with energy
greater than 0.2 $\sqrt{s}$ in the low-angle calorimeter, covering a
polar angle from $1.4^\circ$ to $3.7^\circ$. This criteria corresponds
to retaining quasi-real photons with $\mathrm{\langle Q^2 \rangle
\simeq 0.015 \ {\GeV}^2}$.

About two million photon-photon events are selected by these cuts,
with a background contamination of 0.1\%. Events are further analysed
if they have an identified electron or muon.

Electrons are identified as clusters in the electromagnetic
calorimeter in the polar angular range $|\cos\theta| <0.725$ with
energy above $2\GeV$. They should match a track and have a shower
profile compatible with that expected for an electromagnetic shower.
The ratio $E_{\rm t}/p_{\rm t}$ between the projection of the cluster
energy on the plane transverse to the beams and the transverse momentum
of the track is required to be compatible with unity.  Electrons
due to photon conversion are suppressed by requiring the
distance of closest approach, in the transverse plane,  of the track to the mean $\mathrm{e^{+}
e^{-}}$ collision point in the transverse plane to be less than 0.5 mm
and the invariant mass of the electron candidate and of the closest
track, considered as an electron, to be greater than 0.1 \GeV.

These cuts select 82 events with electron candidates in the 217~pb$^{-1}$
of data collected at $\sqrt{s}=202-209\GeV$, which together with the
137 events previously selected in the data at $\sqrt{s}=189-202\GeV$~\cite{L3bb}
give a total of 219 events with an expected background of 2.0\% from
the $\mathrm{e^{+}e^{-} \ra q \bar{q}}$ and $\mathrm{e^{+}e^{-} \ra
\tau^{+} \tau^{-}}$ processes and a signal efficiency of 1.3\%.

Muon candidates are selected from tracks in the muon spectrometer in
the range $|\cos\theta| <0.8$. A minimal muon momentum of $2\GeV$ is
required to ensure the muons reach the spectrometer after having
crossed the calorimeters. The background from annihilation processes
is suppressed by requiring the muon momentum to be less than
$0.1\sqrt{s}$. Background from cosmic muons is rejected by requiring
the muons to be associated with a signal in the scintillator
time-of-flight system in time with the beam crossing.

After these cuts, 166 events with muon candidates are selected in data
with $\sqrt{s}=202-209\GeV$. Including the 269 events previously
selected at $\sqrt{s}=189-202\GeV$~\cite{L3bb}, a total of 435 events
with muons are retained.  The estimated background from from the
$\mathrm{e^{+}e^{-} \ra q \bar{q}}$, $\mathrm{e^{+}e^{-} \ra \tau^{+}
\tau^{-}}$ and $\mathrm{e^{+}e^{-} \ra e^{+}e^{-} \tau^{+} \tau^{-}}$
is 5.7\% and the signal efficiency is 2.2\%.

\begin{figure}[htbp]
  \begin{center}
    \begin{tabular}{c}
      \mbox{\epsfig{file=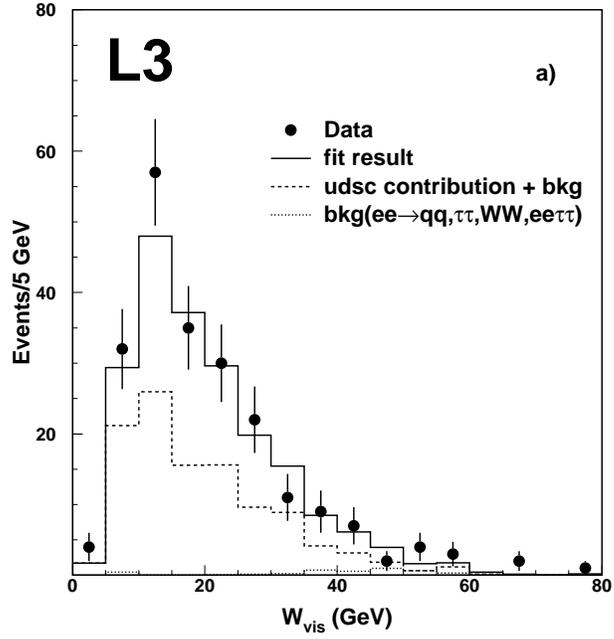, height=.4\textheight}}  \\
      \mbox{\epsfig{file=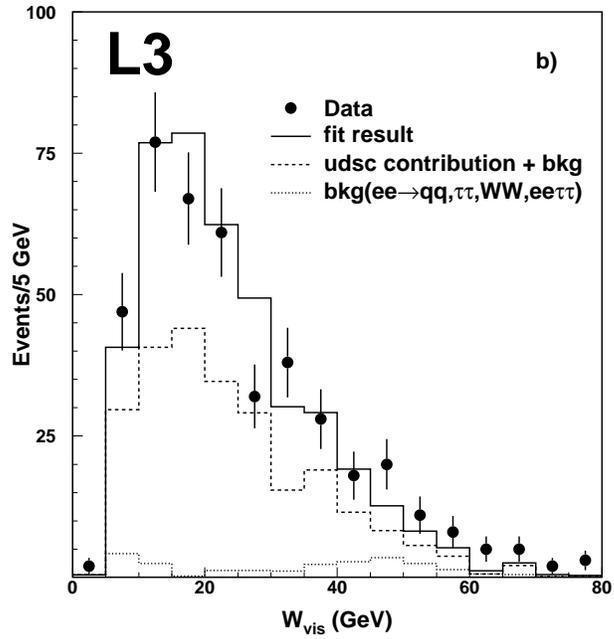, height=.4\textheight}}  \\
    \end{tabular}
  \caption{Visible-mass spectra for the selected $\epem\ra\epem\rm
    hadrons$ events containing (a) an electron or (b) a muon candidate
    at $\sqrt{s}=189-209\GeV$. The points are the data while the
    dotted line represents the background from the $\mathrm{e^{+}e^{-}
    \ra q \bar{q}}$, $\mathrm{e^{+}e^{-} \ra \tau^{+} \tau^{-}}$,
    $\mathrm{e^{+}e^{-} \ra W^{+} W^{-}}$  and 
    $\mathrm{e^{+}e^{-} \ra e^{+}e^{-} \tau^{+} \tau^{-}}$
    processes. The dashed lines are the sum of this background and the
    light-quark contribution, while the solid lines also include
    b-quark production. The normalisation follows from the fit
    discussed in the text.}
    \label{fig:Wvis}
  \end{center}
\end{figure}

Figure~\ref{fig:Wvis} presents the $W_\mathrm{vis}$ spectra of the
selected events for the electron and muon samples.

%%%%%%%%%%%%%%%%%%%%%%%%%%%%%%%%%%%%%%%%%%%%%%%%%%%%%%%%%%%%%%%%%%%%%%%%%%%%%
\section{Results}
%%%%%%%%%%%%%%%%%%%%%%%%%%%%%%%%%%%%%%%%%%%%%%%%%%%%%%%%%%%%%%%%%%%%%%%%%%%%%
 
The cross section for the $\rm\epem\rightarrow\epem b \bar{b} X$
process is determined from the distribution of the transverse momentum
of the lepton with respect to the nearest jet, $P_{\rm t}$. As a
consequence of the large value of $m_{\rm b}$,
the distribution of this variable is enhanced for high values as
compared to the background.  The jets are reconstructed using  
the JADE algorithm~\cite{JADE} with $y_{\rm cut}=0.1$. The  identified
lepton is not included in the
jet. Figure~\ref{fig:ptem_fit} presents the
observed distributions of $P_{\rm t}$ for
electrons and muons.

The data distributions are  fitted using the
least-squares method to the sum of four contributions, whose shapes
are fixed by Monte Carlo 
simulations. The first describes the background
 from annihilation processes and the $\mathrm{e^{+}e^{-} \ra
e^{+}e^{-} \tau^{+} \tau^{-}}$ reaction. Its normalisation,
$N_\mathrm{bkg}$, is fixed to the Monte Carlo predictions listed
in Table~\ref{tab:fit_result}. The  three other contributions are
those from b quarks, c quarks and lighter flavours. Their
normalisations, $N_\mathrm{b\bar{b}}$, $N_\mathrm{c\bar{c}}$ and
$N_\mathrm{uds}$, respectively, are the free parameters of the fit.
The results of the fits are given in Table~\ref{tab:fit_result}: a
b-quark fraction of $46.2 \pm 5.1 \%$ is observed for electrons and
$41.2 \pm 3.8 \%$ for muons, where the uncertainties are statistical.
The $\mathrm{\chi^2}$ per degree of freedom of the fits is acceptable, with
values of $13.7 / 6$ for electrons and $6.4 / 6$ for muons. A
correlation coefficient of about 75\%  between
$N_\mathrm{b\bar{b}}$ and $N_\mathrm{c\bar{c}}$ is observed.
The results of the fits are also graphically shown in Figure~\ref{fig:ptem_fit}.
Figure~\ref{fig:manyplots} presents the distributions of the lepton
momentum, transverse momentum and cosine of polar angle.

The measured fractions of b quarks correspond to observed cross
sections for the luminosity-averaged centre-of-mass
energy $\langle \sqrt{s} \rangle = 198 \GeV$ of:
\begin{eqnarray*}  
\rm\sigma(e^+e^-\rightarrow e^+e^-b \bar b X)^{observed}_{electrons}
& = &   0.41 \pm 0.08 \pm 0.08\rm \ pb\\
\rm\sigma(e^+e^-\rightarrow e^+e^-b \bar b X)^{observed}_{muons}
& = &  0.56 \pm 0.10 \pm 0.10\rm \ pb 
\end{eqnarray*}
The first uncertainties are statistical and the second systematic, and
arise from the sources discussed below. These cross sections
correspond to the phase space of the selected leptons, without any
extrapolation: lepton momenta above $2\GeV$ and polar angles in the
ranges $|\cos\theta|<0.725$ for electrons and $|\cos\theta|<0.8$ for
muons, respectively.

%%%%%%%%%%%%%%%%%%%%%%%%%%%%%%%%%%%%%%%%%%%%%%%%%%%%%%%%%%%%%%%%%%%%%%%%%%%%%
\section{Systematic Uncertainties}
%%%%%%%%%%%%%%%%%%%%%%%%%%%%%%%%%%%%%%%%%%%%%%%%%%%%%%%%%%%%%%%%%%%%%%%%%%%%%

Several potential sources of  systematic uncertainty are considered and
their impact on the observed cross section is detailed in
Table~\ref{tab:syst_bb}.  The 
largest sources of uncertainty arises from the event-selection procedure and
the Monte Carlo modelling of the detector response. Several components
contribute to these uncertainties: the event-selection criteria, the
lepton identification and the detector response and resolution on the
energy and angular variables which identify the fiducial volume. The
effect of these systematic uncertainties is estimated by
varying the corresponding cuts and repeating the fits for the newly
selected event samples. The second most
important source of systematic uncertainty is the jet-reconstruction
method. It is assessed by varying the value of ${y_\mathrm{cut}}$ 
used in the reconstruction of the jets, and performing the fits for
the different $P_{\rm t}$ distributions which are
obtained after the corresponding variation of the jet direction. This
variation also addresses uncertainties in the hadronisation process by
excluding or adding soft clusters to the jets. The
impact of the modelling of c quarks in the event generation is
estimated by repeating fits by using Monte Carlo events generated
with massive matrix elements.  The trigger efficiency is determined
from the data themselves and found to be $(95.6 \pm 2.0) \%$, this uncertainty
is also propagated to the final results. The limited Monte Carlo
statistics has a small impact on the total systematic uncertainty.  In
the fits, the signal events are produced in two separate samples for
the direct and resolved processes and then combined in a 1:1
ratio~\cite{theory}. Systematic uncertainties on this prediction are
estimated by repeating the fits with ratios of 1:2 and 2:1.

%%%%%%%%%%%%%%%%%%%%%%%%%%%%%%%%%%%%%%%%%%%%%%%%%%%%%%%%%%%%%%%%%%%%%%%%%%%%%
\section{Discussion and Conclusions}
%%%%%%%%%%%%%%%%%%%%%%%%%%%%%%%%%%%%%%%%%%%%%%%%%%%%%%%%%%%%%%%%%%%%%%%%%%%%%

The b-production cross sections measured in the phase space of the
selected leptons are compared with the predictions obtained
with the CASCADE Monte Carlo program~\cite{Cascade}. This generator
employs a backward-evolving parton cascade based on the CCFM~\cite{CCFM}
equation. The most important difference as compared to NLO QCD calculations
is the use of an unintegrated gluon density function taking explicitly
into account the transverse momentum distribution of initial state gluons in hard
scattering processes. In NLO QCD, all initial state partons have vanishing
transverse momentum. CASCADE was shown~\cite{HJ} to give a consistent description
of b-quark production at the Tevatron, whereas H1 electro-production
data was found to be in excess by a factor of 2.6. Better agreement was
found with ZEUS electro-production data.

\begin{figure}[htbp]
  \begin{center}
    \begin{tabular}{c}
      \mbox{\epsfig{file=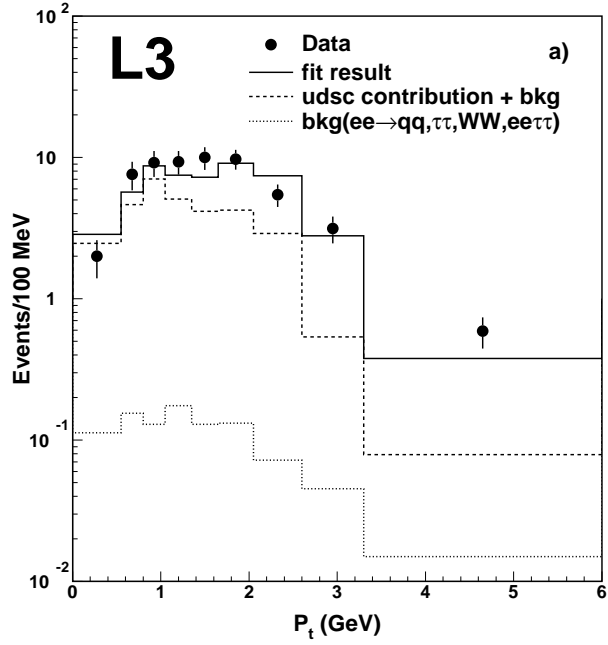, height=.4\textheight}} \\
      \mbox{\epsfig{file=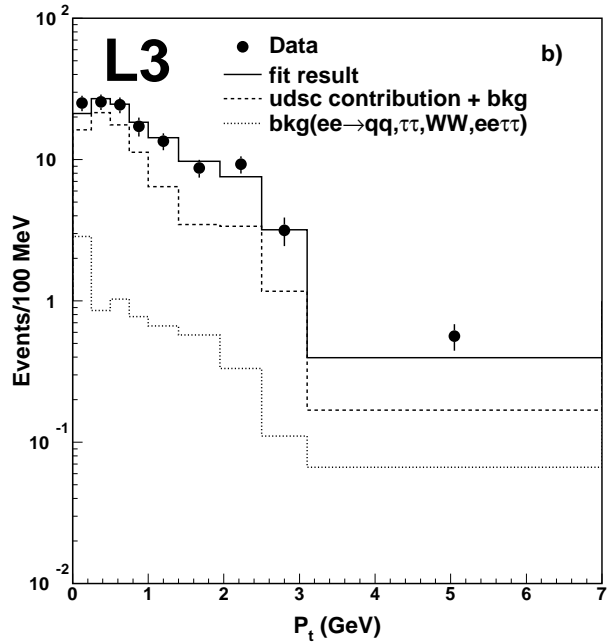, height=.4\textheight}} \\
    \end{tabular}
  \caption{Distributions of the transverse momentum of a) the electron
    candidate and b) the muon candidate with respect to the closest
    jet for the data and the results of the fit. The points are the
    data while the dotted line represents the background from the
    $\mathrm{e^{+}e^{-} \ra q \bar{q}}$, $\mathrm{e^{+}e^{-} \ra
    \tau^{+} \tau^{-}}$, $\mathrm{e^{+}e^{-} \ra W^{+} W^{-}}$ and
    $\mathrm{e^{+}e^{-} \ra e^{+}e^{-} 
    \tau^{+} \tau^{-}}$ processes. The dashed lines are the sum of this
    background and the light-quark contribution, while the solid lines
    also include b-quark production. The normalisation follows from
    the fit discussed in the text.}
  \label{fig:ptem_fit}
  \end{center}
\end{figure}

\begin{figure}[htbp]
  \begin{center}
    \begin{tabular}{cc}
      \mbox{\epsfig{file=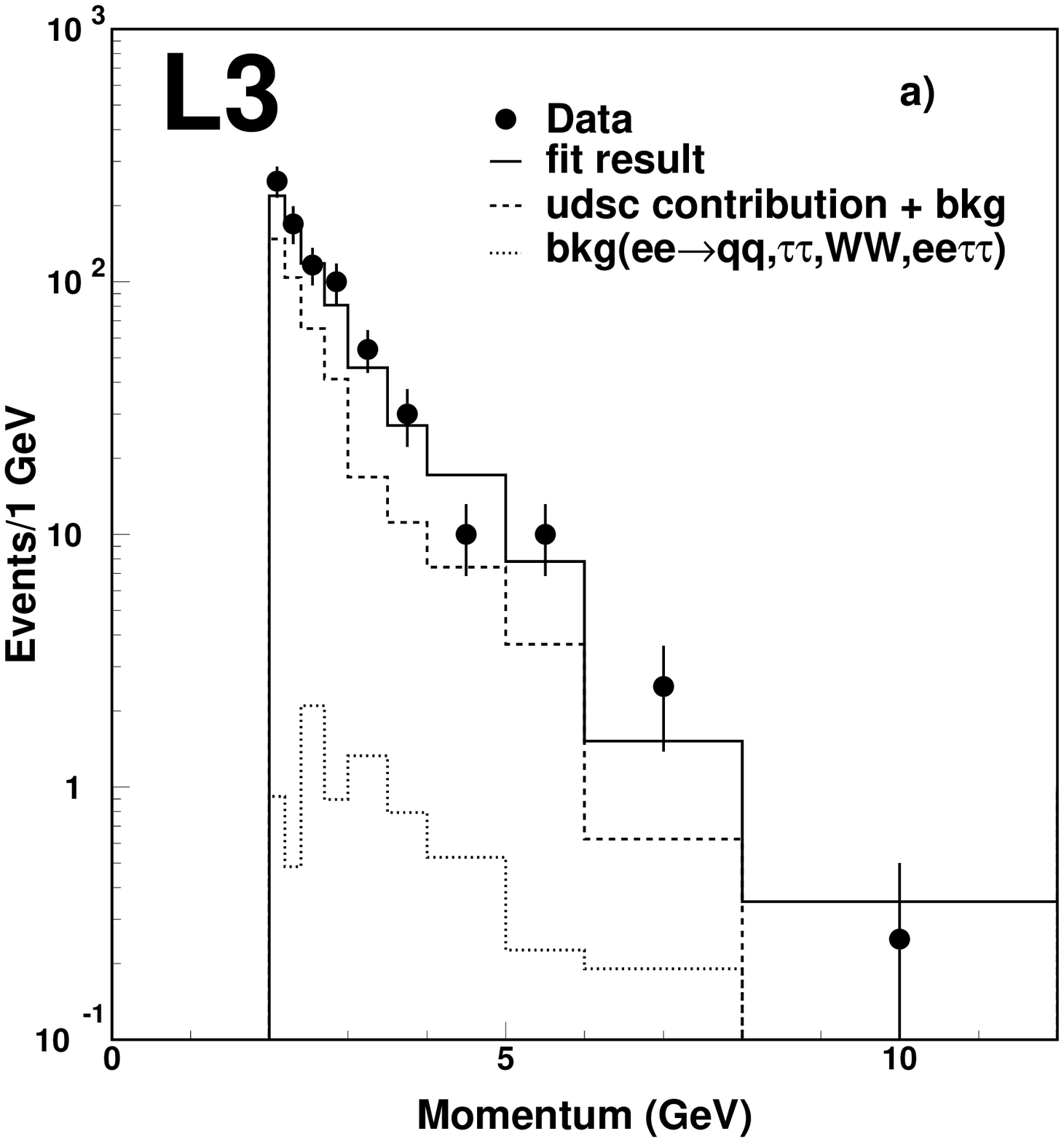, height=.25\textheight}} &
      \mbox{\epsfig{file=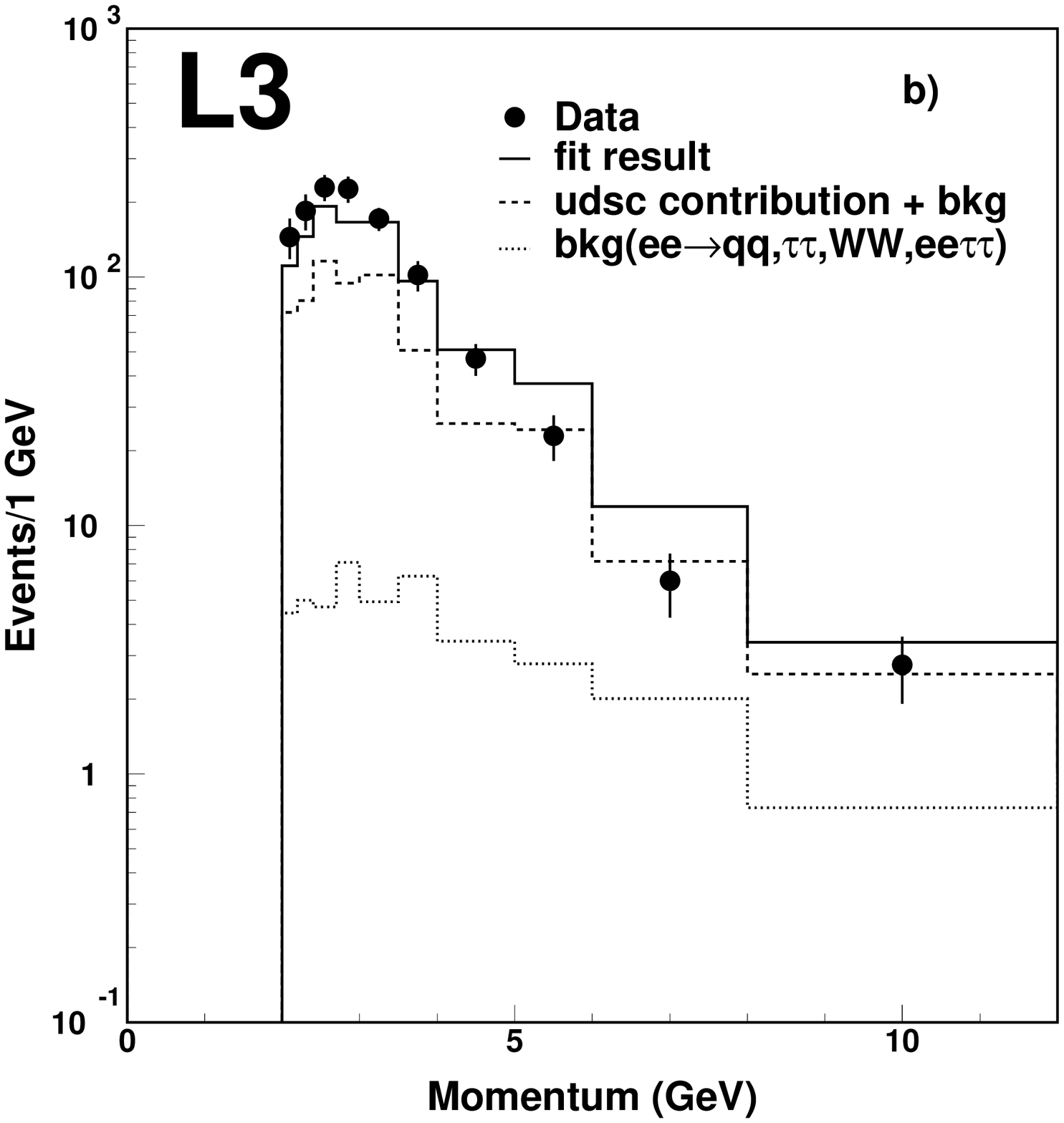, height=.25\textheight}} \\
      \mbox{\epsfig{file=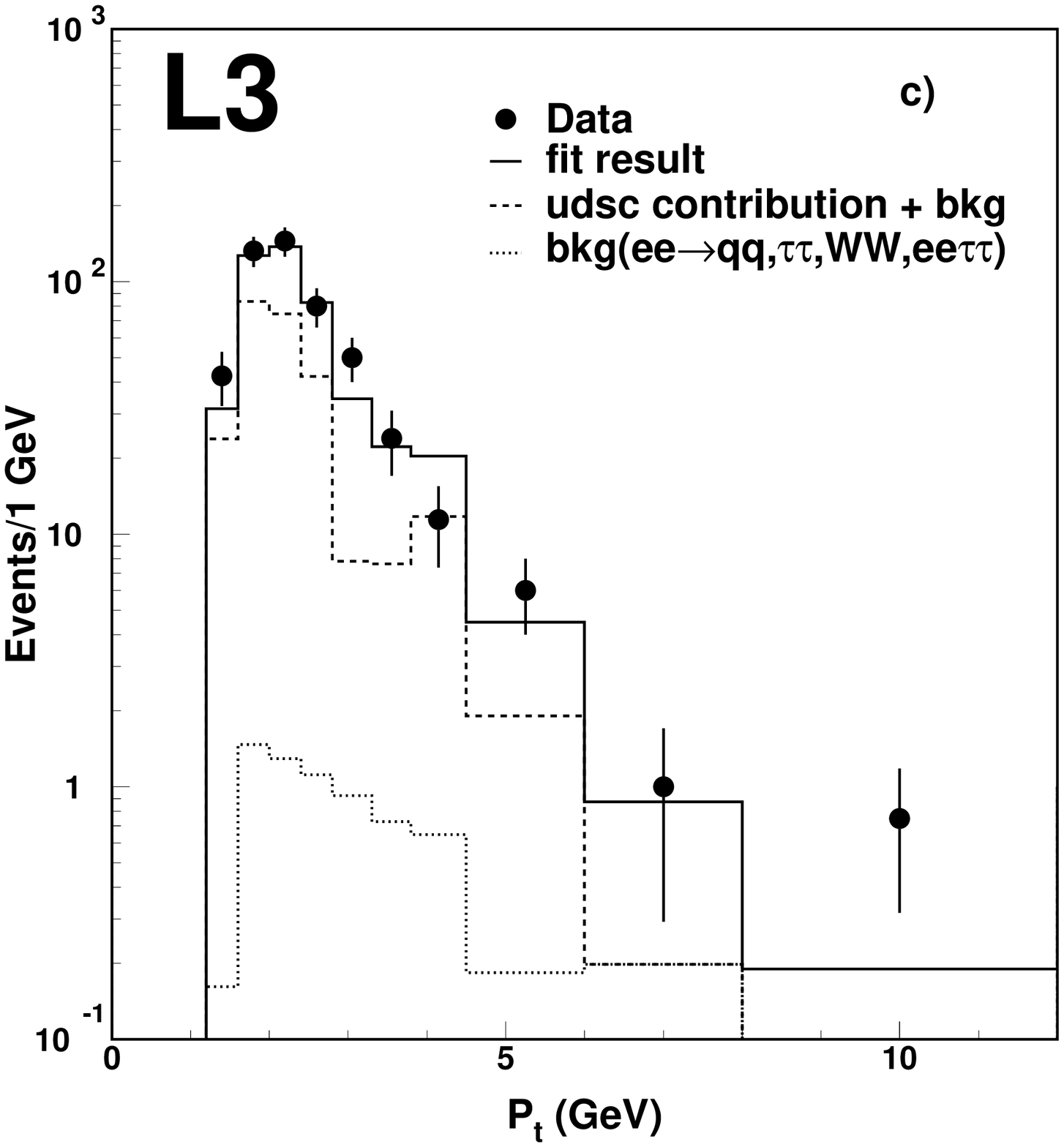, height=.25\textheight}} &
      \mbox{\epsfig{file=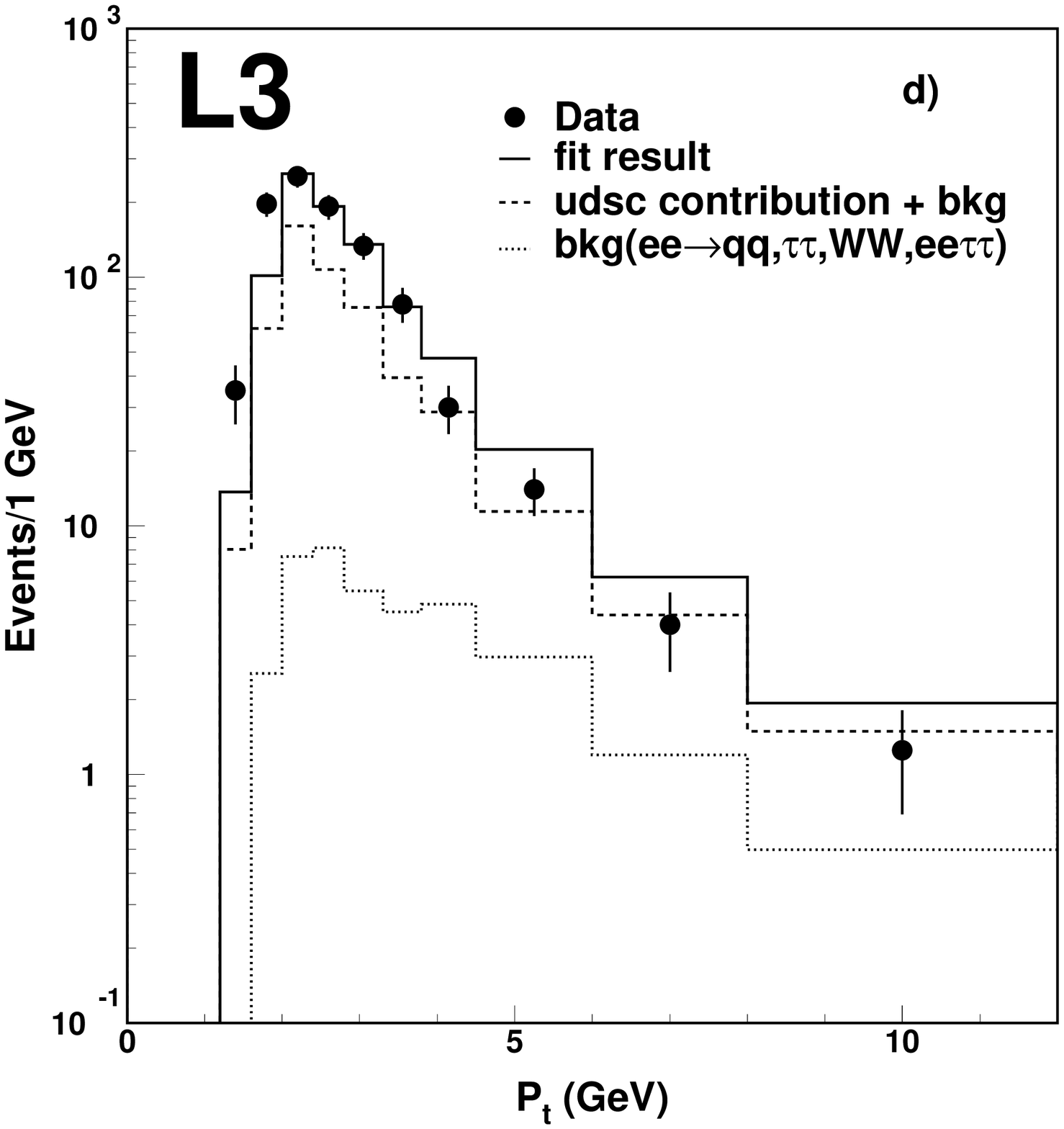, height=.25\textheight}} \\
      \mbox{\epsfig{file=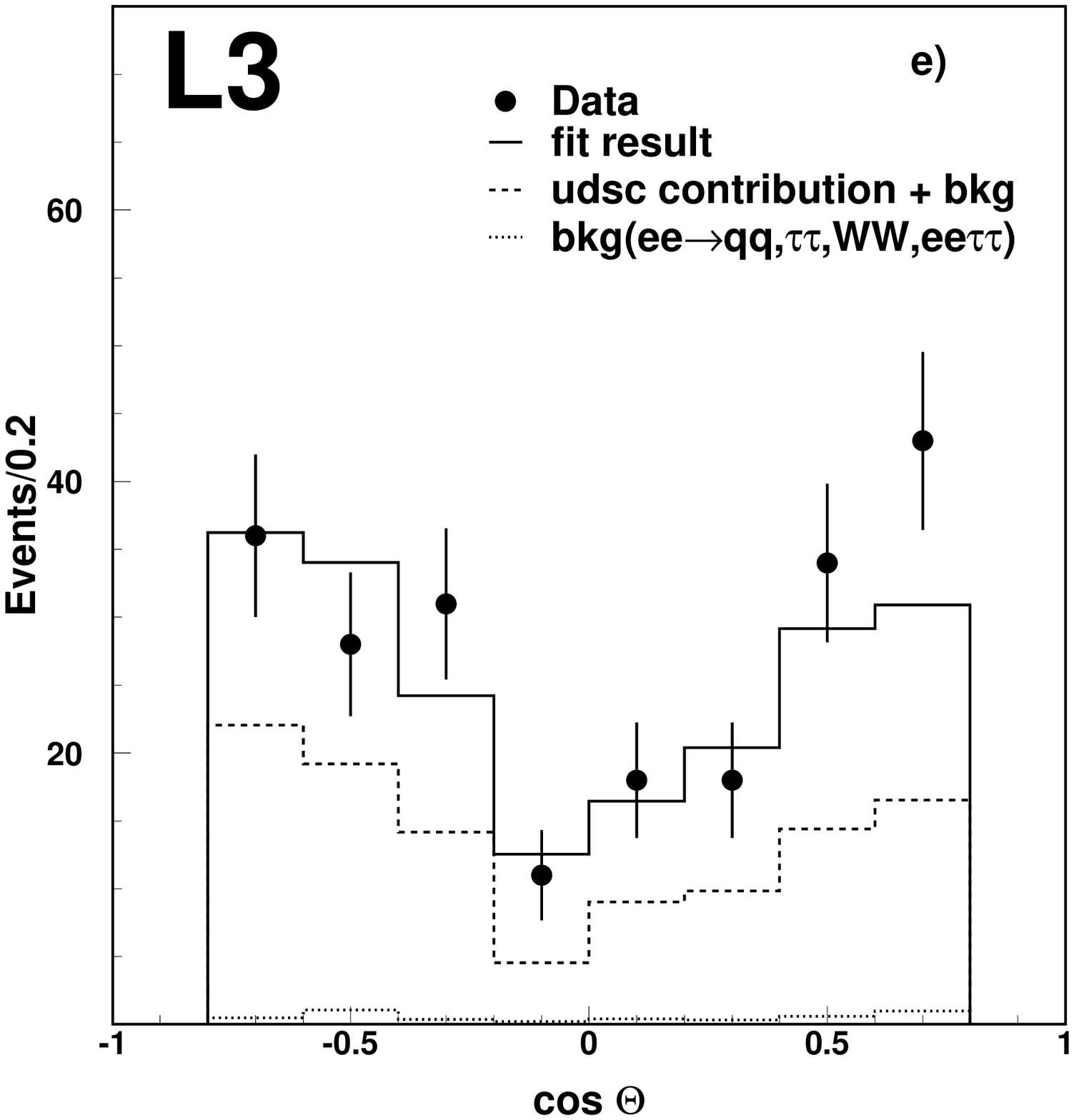, height=.25\textheight}} &
      \mbox{\epsfig{file=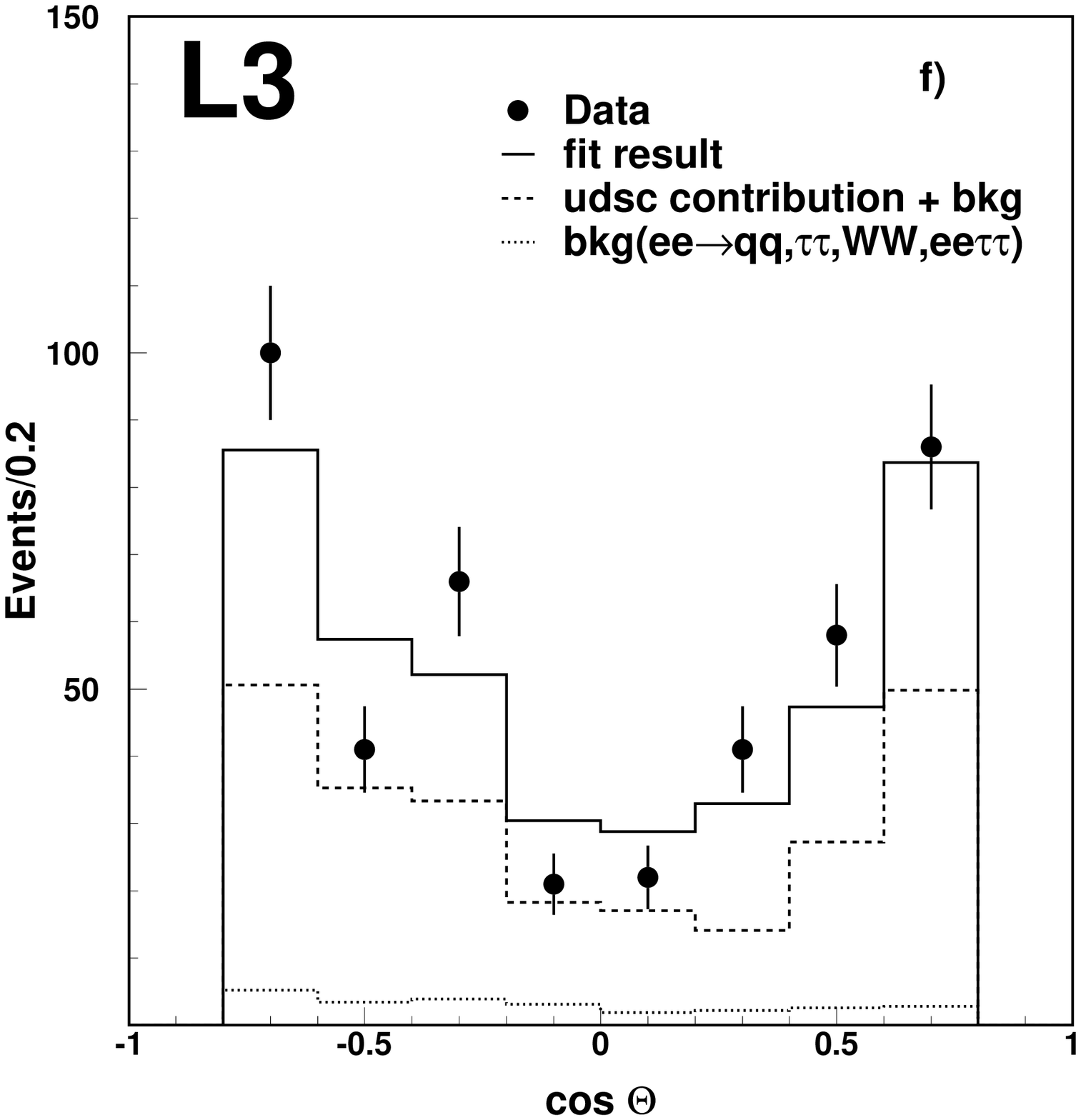, height=.25\textheight}} \\
    \end{tabular}
  \caption{Distribution of a) the lepton momentum, c) its
  transverse momentum and e) the cosine of its  polar angle for
  events containing electrons and b), d) and f) for events containing
  muons. The points are the data while the dotted line
  represents the background from the $\mathrm{e^{+}e^{-} \ra q
  \bar{q}}$, $\mathrm{e^{+}e^{-} \ra \tau^{+} \tau^{-}}$,
  $\mathrm{e^{+}e^{-} \ra W^{+} W^{-}}$ and 
  $\mathrm{e^{+}e^{-} \ra e^{+}e^{-} \tau^{+} \tau^{-}}$
  processes. The dashed lines are the sum of this background and the
  light-quark contribution, while the solid lines also include b-quark
  production. The normalisation follows from the fit discussed in the
  text. }
  \label{fig:manyplots}
  \end{center}
\end{figure}

The comparison of measurements and expectations in the actual phase
space of the selected leptons has the advantage of providing an
assessment of the agreement before any extrapolation is
performed. Summing statistical and systematic uncertainties in
quadrature, one finds:
\begin{displaymath}  
\rm\sigma(e^+e^-\rightarrow e^+e^-b \bar b X)^{observed}_{electrons}
=  0.41 \pm 0.11\ pb\,\,\,\,\,\, 
\rm\sigma(e^+e^-\rightarrow e^+e^-b \bar b X)^{CASCADE}_{electrons}
=  0.11\pm 0.02\ pb
\end{displaymath}
\begin{displaymath}  
\rm\sigma(e^+e^-\rightarrow e^+e^-b \bar b X)^{observed}_{muons}
=  0.56 \pm 0.14\ pb \,\,\,\,\,\, 
\rm\sigma(e^+e^-\rightarrow e^+e^-b \bar b X)^{CASCADE}_{muons}
=  0.14\pm 0.02\ pb 
\end{displaymath}
where the uncertainty on the CASCADE predictions corresponds to a 
 variation of $m_{\rm b}$ in the range $4.75 \pm 0.25\GeV$~\cite{HJMH}.
A disagreement of about three standard deviations is observed for both
flavours of the final-state leptons. This disagreement is mostly due
to the overall normalisation of the sample rather than to a difference
in shape of the most relevant kinematic variables, as also shown in
Figures~\ref{fig:ptem_fit} and~\ref{fig:manyplots}.

The total cross section for open-beauty production in photon-photon
collisions is determined by an extrapolation of the observed cross
section to the full phase space of the process and by correcting for
the semi-leptonic branching ratio of b quarks. The extrapolation
factors are determined with the PYTHIA Monte Carlo program, and
similar results are obtained if the CASCADE Monte Carlo is used. Their
difference, which amounts to 3\%, is considered as an additional
systematic uncertainty. The experimental uncertainties on the
semi-leptonic  branching ratio of b quarks~\cite{branching} is also
propagated to the measurement.

The results for the electron and muon final states read: 
\begin{eqnarray*}
\rm\sigma(e^+e^-\rightarrow e^+e^-b \bar b X)^{total}_{electrons}
& = & 12.6 \pm 2.4 \pm 2.3  \rm \ pb \\
\rm\sigma(e^+e^-\rightarrow e^+e^-b \bar b X)^{total}_{muons}
& = & 13.0 \pm 2.4 \pm 2.3 \rm \ pb, 
\end{eqnarray*}
where the first uncertainty is statistical and the second
systematic. These results are in perfect agreement with each other and
their combination gives:
\begin{displaymath}  
\rm\sigma(e^+e^-\rightarrow e^+e^-b \bar b X)^{total}
=  12.8 \pm 1.7 \pm 2.3\ pb,  
\end{displaymath}
where, again, the first uncertainty is statistical and the second
systematic.  This result is in agreement with our previous measurement
performed with just a subset of the data investigated here~\cite{L3bb}
and has an improved precision.

As a cross check, the values of $N_\mathrm{c\bar{c}}$ found by the fit
are used to extract the total cross section for the production of open charm
at the luminosity-averaged centre-of-mass energy $\langle \sqrt{s}
\rangle = 198 \GeV$ as:
\begin{eqnarray*}
\rm\sigma(e^+e^-\rightarrow e^+e^-c \bar c X)_{electrons}
& = &  (10.4 \pm 1.8) \times 10^2\rm \  pb \\
\rm\sigma(e^+e^-\rightarrow e^+e^-c \bar c X)_{muons}
& = &  \phantom{0}(9.8 \pm 1.6) \times 10^2\rm \  pb, 
\end{eqnarray*}
where uncertainties are statistical. These values agree well, and their average
\begin{displaymath}  
\rm\sigma(e^+e^-\rightarrow e^+e^-c \bar c X)
=  (10.0 \pm 1.2) \times 10^2\  pb
\end{displaymath}
agrees with the dedicated measurement of Reference~\citen{L3bb}, 
$\rm\sigma(e^+e^-\rightarrow e^+e^-c \bar c X)
=  (10.2 \pm 0.3) \times 10^2\  pb$ for  $\langle \sqrt{s}
\rangle = 194 \GeV$, where the uncertainties are statistical only.

An additional cross check showed that values of the open-beauty cross
section determined with the fit procedure discussed above or with a
counting method~\cite{L3bb} are compatible. In the latter case
experimental criteria were chosen to optimise the charm cross section
measurement yielding a result essentially uncorrelated with the
b-quark production rate. 

\begin{figure}[htbp]
  \begin{center}
  \mbox{\epsfig{file=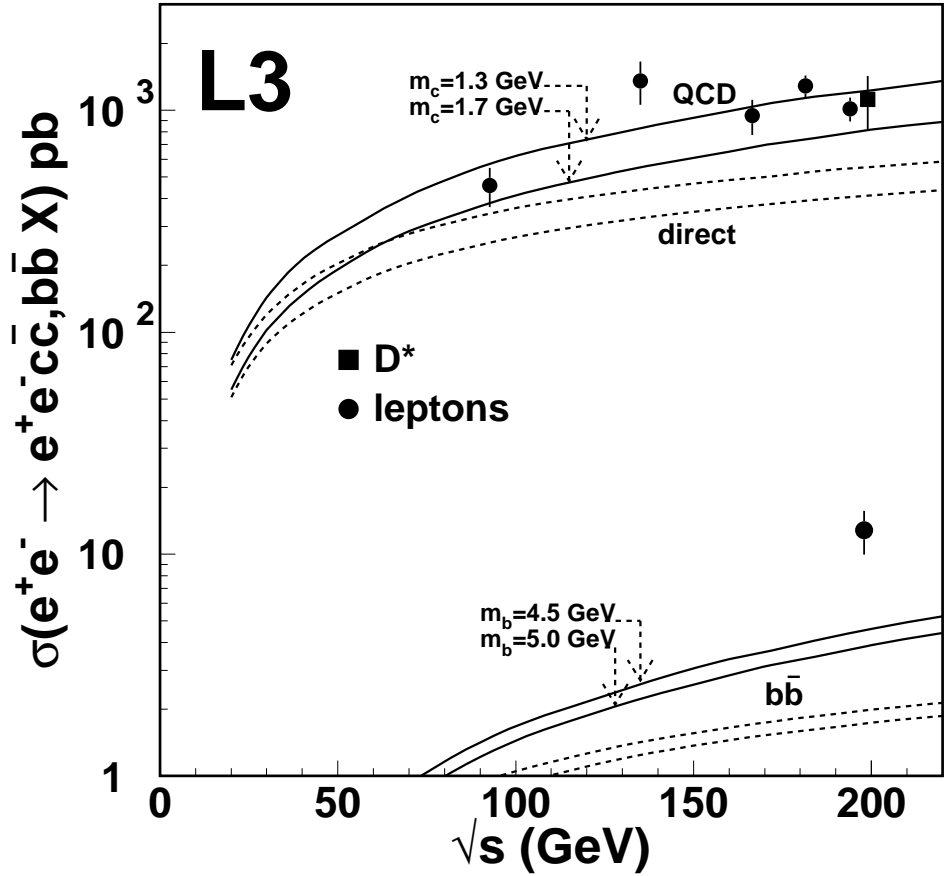,width=.8\textwidth}}
  \caption{The open-charm, upper, and open-beauty, lower, production
    cross sections in photon-photon collisions measured with the L3
    detector. Statistical and systematic uncertainties are added in
    quadrature. The dashed lines correspond to the direct-process
    contribution and the solid lines represent the NLO QCD prediction
    for the sum of the direct and single-resolved processes. The
    effects of a different choice of the values of the quark masses, 
    $m_{\rm c}$ and $m_{\rm b}$, are  illustrated.}
  \label{fig:sigma_ccbb}
  \end{center}
\end{figure}

The total cross section for open-beauty production is compared in
Figure~\ref{fig:sigma_ccbb} to NLO QCD calculations~\cite{theory}.
The dashed line corresponds to the direct process while the solid line
shows the prediction for the sum of direct and resolved processes.
The cross section depends on $m_{\rm b}$, which is varied between $4.5
\GeV$ and $5.0 \GeV$. The threshold for open-beauty production is set
at $10.6 \GeV$.  The theory prediction for the resolved process is
calculated with the GRV parton density function~\cite{GRV}. The same
results are obtained if the Drees-Grassie parton density
function~\cite{DG} is used. For completeness,
Figure~\ref{fig:sigma_ccbb} also compares the cross sections for
open-charm production measured in
References~\citen{L3bb}~and~\citen{L3charm} with the corresponding
predictions.

For $\langle \sqrt{s} \rangle = 198 \GeV$ and $m_{\rm b} = 4.75 \GeV$,
the cross section expected from NLO QCD is $4.1\pm 0.6 $~pb, where the
uncertainty is dominated by uncertainties on the renormalisation scale
and on $m_{\rm b}$. Our measurement is a factor of three, and three
standard deviations, higher than expected. In this respect it is
interesting to remark that the prediction of CASCADE, when
extrapolated to the full phase space, 3.5~pb, agrees with those from
NLO QCD~\cite{HJMH}, and the excess of our data with respect to the
expectations is consistent before and after the extrapolation from the
fiducial volume to the full phase-space.
 
In conclusion, all high-energy data collected by L3 at LEP is
investigated and the $\rm e^+e^-\rightarrow e^+e^-b \bar b X$ cross
sections are measured within the detector fiducial volume and found to
be in excess with respect to Monte Carlo predictions. The cross
sections are extrapolated to the full phase space and found to be in
excess with respect to next-to-leading order QCD calculations. This
confirms our previous findings based on a subset of the full
data-sample.

%%%%%%%%%%%%%%%%%%%%%%%%%%%%%%%%%%%%%%%%%%%%%%%%%%%%%%%%%%%%%%%%%%%%%%%%%%%%%%
% Bibliography
%%%%%%%%%%%%%%%%%%%%%%%%%%%%%%%%%%%%%%%%%%%%%%%%%%%%%%%%%%%%%%%%%%%%%%%%%%%%%%
%

\clearpage
\newpage
%
%%%%%%%%%%%%%%%%%%%%%%%%%%%%%%%%%%%%%%%%%%%%%%%%%%%%%%%%%%%%%%%%%%%%%%%%%%%%%%%
% The author list
%%%%%%%%%%%%%%%%%%%%%%%%%%%%%%%%%%%%%%%%%%%%%%%%%%%%%%%%%%%%%%%%%%%%%%%%%%%%%%%
%

\typeout{   }     
\typeout{Using author list for paper 287 -  }
\typeout{$Modified: Jul 15 2001 by smele $}
\typeout{!!!!  This should only be used with document option a4p!!!!}
\typeout{   }
%
%
%
%  L A T E X  version!!
%
%
% Make sure that the Lep package has been used!
%\input{Lep.sty}%
%
%\ifx\LepCalled\undefined%
%\typeout{     }%
%\typeout{!!!!!!!!!!!!!!!!!!!!!!!!!!!!!!!!!!!!!!!!!!!!!!!!!!!!!!!!!!!}%
%\typeout{Yikes.  You haven't used the Lep package!}%
%\typeout{Please put \protect\usepackage\protect{Lep\protect} in your preamble,
%         followed by}%
%\typeout{\protect\Lep\protect{1\protect} or \protect\Lep\protect{2\protect}}%
%\typeout{     }%
%\typeout{For now you will get a Lep phase 2 authorlist (may not be right!).}%
%\typeout{!!!!!!!!!!!!!!!!!!!!!!!!!!!!!!!!!!!!!!!!!!!!!!!!!!!!!!!!!!!}%
%\typeout{     }%
%\Lep{2}\fi%

\newcount\tutecount  \tutecount=0
\def\tutenum#1{\global\advance\tutecount by 1 \xdef#1{\the\tutecount}}
\def\tute#1{$^{#1}$}
\tutenum\aachen            % 1 
\tutenum\nikhef            % 2 
\tutenum\mich              % 3 
\tutenum\lapp              % 4 
\tutenum\basel             % 5 
\tutenum\lsu               % 6 
\tutenum\beijing           % 7 
\tutenum\bologna           % 8 
\tutenum\tata              % 9 
\tutenum\ne                % 10
\tutenum\bucharest         % 11
\tutenum\budapest          % 12
\tutenum\mit               % 13
\tutenum\panjab            % 14 
\tutenum\debrecen          % 15
\tutenum\dublin            % 16
\tutenum\florence          % 17
\tutenum\cern              % 18
\tutenum\wl                % 19
\tutenum\geneva            % 20
\tutenum\hamburg           % 21
\tutenum\hefei             % 22
\tutenum\lausanne          % 23
\tutenum\lyon              % 24
\tutenum\madrid            % 25
\tutenum\florida           % 26
\tutenum\milan             % 27
\tutenum\moscow            % 29
\tutenum\naples            % 30
\tutenum\cyprus            % 31
\tutenum\nymegen           % 32
\tutenum\caltech           % 33
\tutenum\perugia           % 34
\tutenum\peters            % 35
\tutenum\cmu               % 36
\tutenum\potenza           % 37
\tutenum\prince            % 38
\tutenum\riverside         % 39
\tutenum\rome              % 40
\tutenum\salerno           % 41
\tutenum\ucsd              % 42
\tutenum\sofia             % 43
\tutenum\korea             % 44
\tutenum\taiwan            % 45
\tutenum\tsinghua          % 46
\tutenum\purdue            % 47
\tutenum\psinst            % 48
\tutenum\zeuthen           % 49
\tutenum\eth               % 50

{
\parskip=0pt
\noindent
{\bf The L3 Collaboration:}
\ifx\selectfont\undefined%  old style font selection
 \baselineskip=10.8pt
 \baselineskip\baselinestretch\baselineskip
 \normalbaselineskip\baselineskip
 \ixpt
\else%                      new style font selection
 \fontsize{9}{10.8pt}\selectfont
\fi
\medskip
\tolerance=10000
\hbadness=5000
\raggedright
\hsize=162truemm\hoffset=0mm
\def\r{\rlap,}
\noindent

P.Achard\r\tute\geneva\ 
O.Adriani\r\tute{\florence}\ 
M.Aguilar-Benitez\r\tute\madrid\ 
J.Alcaraz\r\tute{\madrid}\ 
G.Alemanni\r\tute\lausanne\
J.Allaby\r\tute\cern\
A.Aloisio\r\tute\naples\ 
M.G.Alviggi\r\tute\naples\
H.Anderhub\r\tute\eth\ 
V.P.Andreev\r\tute{\lsu,\peters}\
% ...
\newpage
%\rule{\textwidth}{0.4pt}
\begin{list}{A}{\itemsep=0pt plus 0pt minus 0pt\parsep=0pt plus 0pt minus 0pt
                \topsep=0pt plus 0pt minus 0pt}
\item[\aachen]
 III. Physikalisches Institut, RWTH, D-52056 Aachen, Germany$^{\S}$
\item[\nikhef] National Institute for High Energy Physics, NIKHEF, 
     and University of Amsterdam, NL-1009 DB Amsterdam, The Netherlands
\item[\mich] University of Michigan, Ann Arbor, MI 48109, USA
\item[\lapp] Laboratoire d'Annecy-le-Vieux de Physique des Particules, 
     LAPP,IN2P3-CNRS, BP 110, F-74941 Annecy-le-Vieux CEDEX, France
\item[\basel] Institute of Physics, University of Basel, CH-4056 Basel,
     Switzerland
\item[\lsu] Louisiana State University, Baton Rouge, LA 70803, USA
\item[\beijing] Institute of High Energy Physics, IHEP, 
  100039 Beijing, China$^{\triangle}$ 
\item[\bologna] University of Bologna and INFN-Sezione di Bologna, 
     I-40126 Bologna, Italy
\item[\tata] Tata Institute of Fundamental Research, Mumbai (Bombay) 400 005, India
\item[\ne] Northeastern University, Boston, MA 02115, USA
\item[\bucharest] Institute of Atomic Physics and University of Bucharest,
     R-76900 Bucharest, Romania
\item[\budapest] Central Research Institute for Physics of the 
     Hungarian Academy of Sciences, H-1525 Budapest 114, Hungary$^{\ddag}$
\item[\mit] Massachusetts Institute of Technology, Cambridge, MA 02139, USA
\item[\panjab] Panjab University, Chandigarh 160 014, India
\item[\debrecen] KLTE-ATOMKI, H-4010 Debrecen, Hungary$^\P$
\item[\dublin] Department of Experimental Physics,
  University College Dublin, Belfield, Dublin 4, Ireland
\item[\florence] INFN Sezione di Firenze and University of Florence, 
     I-50125 Florence, Italy
\item[\cern] European Laboratory for Particle Physics, CERN, 
     CH-1211 Geneva 23, Switzerland
\item[\wl] World Laboratory, FBLJA  Project, CH-1211 Geneva 23, Switzerland
\item[\geneva] University of Geneva, CH-1211 Geneva 4, Switzerland
\item[\hamburg] University of Hamburg, D-22761 Hamburg, Germany
\item[\hefei] Chinese University of Science and Technology, USTC,
      Hefei, Anhui 230 029, China$^{\triangle}$
\item[\lausanne] University of Lausanne, CH-1015 Lausanne, Switzerland
\item[\lyon] Institut de Physique Nucl\'eaire de Lyon, 
     IN2P3-CNRS,Universit\'e Claude Bernard, 
     F-69622 Villeurbanne, France
\item[\madrid] Centro de Investigaciones Energ{\'e}ticas, 
     Medioambientales y Tecnol\'ogicas, CIEMAT, E-28040 Madrid,
     Spain${\flat}$ 
\item[\florida] Florida Institute of Technology, Melbourne, FL 32901, USA
\item[\milan] INFN-Sezione di Milano, I-20133 Milan, Italy
\item[\moscow] Institute of Theoretical and Experimental Physics, ITEP, 
     Moscow, Russia
\item[\naples] INFN-Sezione di Napoli and University of Naples, 
     I-80125 Naples, Italy
\item[\cyprus] Department of Physics, University of Cyprus,
     Nicosia, Cyprus
\item[\nymegen] Radboud University and NIKHEF, 
     NL-6525 ED Nijmegen, The Netherlands
\item[\caltech] California Institute of Technology, Pasadena, CA 91125, USA
\item[\perugia] INFN-Sezione di Perugia and Universit\`a Degli 
     Studi di Perugia, I-06100 Perugia, Italy   
\item[\peters] Nuclear Physics Institute, St. Petersburg, Russia
\item[\cmu] Carnegie Mellon University, Pittsburgh, PA 15213, USA
\item[\potenza] INFN-Sezione di Napoli and University of Potenza, 
     I-85100 Potenza, Italy
\item[\prince] Princeton University, Princeton, NJ 08544, USA
\item[\riverside] University of Californa, Riverside, CA 92521, USA
\item[\rome] INFN-Sezione di Roma and University of Rome, ``La Sapienza",
     I-00185 Rome, Italy
\item[\salerno] University and INFN, Salerno, I-84100 Salerno, Italy
\item[\ucsd] University of California, San Diego, CA 92093, USA
\item[\sofia] Bulgarian Academy of Sciences, Central Lab.~of 
     Mechatronics and Instrumentation, BU-1113 Sofia, Bulgaria
\item[\korea]  The Center for High Energy Physics, 
     Kyungpook National University, 702-701 Taegu, Republic of Korea
\item[\taiwan] National Central University, Chung-Li, Taiwan, China
\item[\tsinghua] Department of Physics, National Tsing Hua University,
      Taiwan, China
\item[\purdue] Purdue University, West Lafayette, IN 47907, USA
\item[\psinst] Paul Scherrer Institut, PSI, CH-5232 Villigen, Switzerland
\item[\zeuthen] DESY, D-15738 Zeuthen, Germany
\item[\eth] Eidgen\"ossische Technische Hochschule, ETH Z\"urich,
     CH-8093 Z\"urich, Switzerland
\item[\S]  Supported by the German Bundesministerium 
        f\"ur Bildung, Wissenschaft, Forschung und Technologie.
\item[\ddag] Supported by the Hungarian OTKA fund under contract
numbers T019181, F023259 and T037350.
\item[\P] Also supported by the Hungarian OTKA fund under contract
  number T026178.
\item[$\flat$] Supported also by the Comisi\'on Interministerial de Ciencia y 
        Tecnolog{\'\i}a.
\item[$\sharp$] Also supported by CONICET and Universidad Nacional de La Plata,
        CC 67, 1900 La Plata, Argentina.
\item[$\triangle$] Supported by the National Natural Science
  Foundation of China.
\end{list}
}
\vfill

%%% Local Variables: 
%%% mode: latex
%%% TeX-master: t
%%% End:

\newpage

%%%%%%%%%%%%%%%%%%%%%%%%%%%%%%%%%%%%%%%%%%%%%%%%%%%%%%%%%%%%%%%%%%%%%%%%%%%%%%
% Tables
%%%%%%%%%%%%%%%%%%%%%%%%%%%%%%%%%%%%%%%%%%%%%%%%%%%%%%%%%%%%%%%%%%%%%%%%%%%%%%%

\begin{table}[htb]
  \begin{center}
    \begin{tabular}{|l|c|c|} 
      \cline{2-3}
      \multicolumn{1}{l|}{}          & Electrons                       & Muons                              \\ \hline
      ${N_\mathrm{bkg}}$             & 4.4 (fixed)                     & 24.8 (fixed)                       \\ 
      ${N_\mathrm{b\bar{b}}}$        & $\phantom{0}94.3 \pm 18.3$      & $172.0 \pm 31.0$                   \\
      ${N_\mathrm{c\bar{c}}}$        & $105.4 \pm 17.9$                & $220.5 \pm 35.4$                   \\ 
      ${N_\mathrm{uds}}$             & $0.0^{+12.0}_{-\phantom{0}0.0}$ & $0.0^{ + 52.3}_{-\phantom{0}0.0}$  \\ 
      $\mathrm{\chi^2}$ / d.o.f.     & 13.7 / 6                        & 6.4 / 6                            \\  \hline 

    \end{tabular}
  \end{center}
  \caption{Results of the fit to the distribution of 
the transverse momentum of the lepton with respect to the
nearest jet. The fit parameters are constrained to be positive. The correlation between $\mathrm{N_{b\bar{b}}}$  and $\mathrm{N_{c\bar{c}}}$ is 75\%.}
  \label{tab:fit_result} 
\end{table}

\begin{table}[htb]
  \begin{center}
    \begin{tabular}{|l|c|c|} 
      \cline{2-3}
      \multicolumn{1}{l|}{}        & \multicolumn{2}{c|}{Uncertainty on cross section (\%)}\\
      \hline
       Source of uncertainty         & Electrons & Muons\\
      \hline
      Event selection                    & \phantom{0}6.0 & 10.4\\ 
      Lepton identification              & \phantom{0}7.9 & \phantom{0}2.2 \\
      Fiducial volume                    &           12.3 & 10.0 \\
      Jet reconstruction                 & \phantom{0}8.2 & \phantom{0}8.2 \\
      Massive/massless charm             & \phantom{0}3.0 & \phantom{0}3.0 \\
      Trigger efficiency                 & \phantom{0}2.0 & \phantom{0}2.0 \\ 
      Monte Carlo statistics             & \phantom{0}1.6 & \phantom{0}1.4 \\ 
      Direct / resolved ratio            & \phantom{0}0.1 & \phantom{0}1.0 \\ 
      \hline                                                              
      Total                              & 18.3           & 17.2\\ 
      \hline
    \end{tabular}
  \end{center}
  \caption{Systematic uncertainties on the observed values
of the cross section of the process $\mathrm{e^+e^-\rightarrow e^+e^-b
\bar b X}$ for events tagged by electrons or muons. An additional
uncertainty of 3\% affects the extrapolation to the total cross section.}
  \label{tab:syst_bb} 
\end{table}

%%%%%%%%%%%%%%%%%%%%%%%%%%%%%%%%%%%%%%%%%%%%%%%%%%%%%%%%%%%%%%%%%%%%%%%%%%%%%%
% Figures
%%%%%%%%%%%%%%%%%%%%%%%%%%%%%%%%%%%%%%%%%%%%%%%%%%%%%%%%%%%%%%%%%%%%%%%%%%%%%%%

\newpage

\newpage

\newpage

\newpage

\newpage

\end{document}